\documentclass[10pt,thesis]{article}
\usepackage{graphicx}
\usepackage{amsmath}
\usepackage{rotating}
\usepackage{lipsum}
\usepackage{graphicx}
\usepackage{multicol}
\usepackage{hyperref}
\usepackage{amsmath}
\usepackage{amsfonts}
\usepackage{setspace}
\usepackage[margin=3.2cm]{geometry}
\usepackage{graphicx}
\usepackage{capt-of}
\usepackage{courier}
\usepackage{color}
\usepackage[usenames,dvipsnames]{xcolor}
\usepackage{fancyhdr}

\begin{document}

\title{A survey on Human Mobility and its Applications}
\author{Fereshteh Asgari$^\star$, Vincent Gauthier$^\star$, Monique Becker$^\star$ 
\let\thefootnote\relax\footnote{Lab. CNRS SAMOVAR UMR 5157, Telecom SudParis, Evry, France}  \\
	}
\date{June 2013}
\maketitle

\newcommand*{\titleAT}{\begingroup 
\newlength{\drop} 
\drop=0.1\textheight 

\rule{\textwidth}{1pt}\par 
\vspace{2pt}\vspace{-\baselineskip} 
\rule{\textwidth}{0.4pt}\par 

\vspace{\drop} 
\centering 
\textcolor{blue}{ 
{\Huge State of The Art}\\[0.5\baselineskip] 
{\Large on}\\[0.75\baselineskip] 
{\Huge HUMAN MOBILITY}} 

\vspace{0.25\drop} 
\rule{0.3\textwidth}{0.4pt}\par 
\vspace{\drop} 

{\Large \textsc{Fereshteh ASGARI, Vincent Gauthier, Monique Becker}}\par 

\vfill 
{ \textsc{Telecom Sudparis}}\par 

\vspace*{\drop} 

\rule{\textwidth}{0.4pt}\par 
\vspace{2pt}\vspace{-\baselineskip} 
\rule{\textwidth}{1pt}\par 

\endgroup}

\section*{Abstract}

Human Mobility has attracted attentions from different fields of studies such as epidemic modeling, traffic engineering, traffic prediction and  urban planning.
In this survey we review major characteristics of human mobility studies including from trajectory-based studies to studies using graph and network theory.
In trajectory-based studies statistical measures such as jump length distribution and radius of gyration are analyzed in order to investigate how people move in their daily life, and if it is possible to model this individual movements and make prediction based on them.
Using graph in mobility studies, helps to investigate the dynamic behavior of the system, such as diffusion and flow in the network and makes it easier to estimate how much one part of the network influences another by using metrics like centrality measures. 
We aim to study population flow in transportation networks using mobility data to derive models and patterns, and to develop new applications in predicting phenomena such as congestion.
Human Mobility studies with the new generation of mobility data provided by cellular phone networks, arise new challenges such as data storing, data representation, data analysis and computation complexity. 
A comparative review of different data types used in current tools and applications of Human Mobility studies leads us to new approaches for dealing with mentioned challenges.

\begin{multicols}{2}
\section{Introduction}
Knowing the flows of individuals from one point to the other in a city or a country provides us with useful information for modeling Human Mobility behaviors and characteristics which could be used from different aspects. \\
First, Human Mobility is related to the fundamental problem in geography and spatial economics of the location of activities and their spatial distribution.
An important example is traffic systems: Analyzing huge amount of mobility data, one purpose is 
to study and model traffic flow in road networks and public transportation networks. Another purpose is to be able to predict the flow among these networks and possibly to predict the future position of a moving object (individual or a vehicle).\\
Second, Human Mobility is a key factor in spreading of infectious diseases. These diseases transmit between humans in close proximity/transmission area and contaminate the population because people travel and interact.
It has been shown that, human mobility exhibit a set of scaling relations and statistical regularities. Thus, the studies on how the topological features of traffic networks can be incorporated in models for disease dynamics have become interesting. It has been observed that the way topology is translated into dynamics can have a profound impact on the overall disease dynamics. 
At a global level for example, the spreading of infectious diseases such as the flu is - in addition to seasonal effects - largely controlled by air travel patterns.\\
Third, statistics about individual movements are important for more commercial applications such as geomarketing. Finding the hot spot to place your advertisement depends on the number of people going through location location and thus implies to know the flows, or at least the dominant flows of individuals in the area under consideration. Recommendation system relying on region  of interest \cite{RoI} of population is another example \cite{Barth}
\cite{hist}.\\ 
According to Basol \cite{valImp} , the value of mobility reaches far beyond mere geographical movement of humans, but provides a complete new mindset on human interactions which could be considered from spatial, temporal, and contextual aspects.
Different dimensions of Human mobility also have been discussed in Kakihara and Sorensen's study \cite{mAspects}. They have mentioned the importance of "being mobile" not just as a matter of people traveling  but,  related to  the  interaction they perform -- the way in which they interact with each other in  their  social  lives.
 Considering this importance, they have expanded  the  concept  of mobility by looking  at three  distinct  dimensions;  namely, spatial, temporal and contextual  mobility. 
 They have also discussed about  issues related to \emph{virtual community} or \emph{cyber community} \cite{mAspects} which today is known as social network.
    Karamshuk et al \cite{Oppnet} have developed the idea of different aspects of Human Mobility introduced in \cite{valImp} and presented the properties of Human Mobility in three main different dimensions : spatial, temporal and social aspects.
Each of these aspects also has been studied in different scales; as a spatial network, human mobility have been studied in global , continent scale \cite{multiscale}, country scale \cite{multiscale} , \cite{PCA} , regional scale \cite{China} , city scale \cite{Hunter2011}, \cite{UnveilComplex} and much finer scales such as campus or building scale \cite{compNetAnHuM} \cite{empS}. 
However, human mobility occurs on a variety of  length scales, ranging from short distances to long-range travel by air, and involves diverse methods of transportation (public transportation, roads, highways, trains, and air transportation).
No comprehensive study exists that incorporates traffic on all spatial scales  \cite{NonDy}. This would require the collection and compilation of data for various transportation networks into a multi-component data set; a difficult task particularly on an international scale \cite{NonDy} (e.g. \cite{vespignani}). Finding the proper scale for Mobility studies has been discussed in some studies \cite{optReso},\cite{PCA}, \cite{Sptmp}.
The authors in \cite{optReso} have developed discussion about the scale of spatial network in human mobility studies. They try to study if there is an optimal spatial resolution for the analysis of Human Mobility. They  build a multi-resolution grid and map the trajectories with several complex networks, by connecting the different areas of region of interest. Then they analyze the structural properties of these networks and the quality of the boarders it is possible to infer from them.\\
Regarding temporal aspect of Human Mobility, there are also some studies \cite{PCA}, \cite{Sptmp}, 
 that have been trying to investigate the periodic patterns of human mobility 
with extracting daily and weekly periodic patterns recognized in mobility data.
The importance of this aspect of human mobility raised when we want to study the dynamic of Human Mobility and how it changes by time. Different time intervals could be considered for these studies: from long-term such as monthly intervals to short-term such as hourly or even finer time intervals.  The length of time interval in dynamic analysis should be chosen such that enough events are collected for any measures to be meaningful \cite{path}, in another word, the $\Delta t$ should be small enough to give meaningful results for our purpose. Regarding the importance of temporal variation in Human Mobility studies the authors in \cite{perPerio} illustrate that incorrect choosing of $\Delta t$ in snapshots, may impose a strong bias on the resulting analysis and conclusion. The importance of choosing proper time interval is a concern in both data collecting and analysis aspect.\\
Beside significant interests in mobility studies, the explosive growth in mobile devices, the emergence and convergence of information and communication technologies telecommunication facilities provide us with a variety of recorded data of human mobility and social communication which is a valuable source of knowledge. This data depending on collection technologies have different properties which require new approaches of analyzing. 

In the rest of this report we present different data collection techniques for Human Mobility studies. Then we present a summary of studies on Human Mobility.
In Human Mobility studies, we have distinguished three different baselines which could help us to categorize these studies. However there are overlaps in these area of studies and we can not say they are totally separated. 
We continue with the a review on tools and applications in Human Mobility studies and the report will be finished with discussions on new challenges in Human Mobility studies and some perspective on these issues.

\section{Related works}
Human Mobility studies try to answer a wide range of questions based on different aspects of human mobility. 
A considerable amount of Human Mobility studies are trajectory-based studies in which individuals trajectories are traced and  the behavior of these trajectories are studied. These studies 
are trying to answer the following questions:
How far do people travel every day? \cite{barab}
What are the main measures in Human Mobility studies?
How these measures represent mobility behaviors of individuals?
Does human mobility follows any model or pattern? \cite{barab}, \cite{uniModMob}, \cite{tale}
Is it possible to estimate the trajectory due to home-to-work commutes?
If the trajectories pattern depends on Geographical position of individuals? \cite{China}
How different metropolitan areas exhibit distinct mobility patterns due to differences in geographic distributions of homes and jobs, transportation infrastructures, and other factors? \cite{HMmetro}
Is it possible to predict the next position of individuals having previous records of their trajectories? \cite{barab} \cite{vtrack} \cite{ctrack} \cite{Hunter2011}\\
 
\noindent The main focus of these approaches is spatial characteristics (measures) of movements and how they change in Human Mobility. Measures such as jump length, center of the mass and radius of gyration are the basic of these studies \cite{barab}. 
The trajectory data set in these studies, mainly GPS data or CDRs.  \\
Also studying statistical characteristics and patterns of human movements have revealed a number of scaling
properties in human trajectories: Gonzalez et al in \cite{barab} and Brockmann et al \cite{brock} showed a truncated power-law tendency in the distribution of jump length. \\
It was observed that most individuals travel only over short distance, but a few regularly more over hundred kilometers. Further studies \cite{uniModMob} \cite{barab} showed that the individual travel patterns collapse into a single spatial probability distribution, indicating that, despite the diversity of their travel history, humans follow simple reproducible patterns.\\ 
More precisely, a ``trajectory'' can be defined \cite{trajPatMin} as a spatio-temporal sequence of triples $(x_i,y_i,t_i)$, where $t_i$ is a time stamp and $\overrightarrow{r_{i}^a}=(x_i,y_i)$ are points in $\mathbb{R}^2$. 
Having the sequence of consecutive traces, jump length distribution is an important measure that have been investigating in Human Mobility. 
Brockman et al \cite{brock}, analyzed a huge data set of records of bank notes circulation, interpreting them as a proxy of human movements \cite{Oppnet}, showed that travel distances $\Delta r$  of individuals follow a power-law distribution  \[ P(\Delta r) \sim (\Delta r)^ {–(1+\beta)} \] where $\beta <2 $.
 This fits the intuition that people usually move over short distances, whereas occasionally they take rather long trips. The distribution known as Levy Flight, was previously observed 
  as an approximation of migration trajectories among different animal species. 
Studying data collected tracing mobile phone users, Gonzalez et al \cite{barab} complemented the previous finding with an exponential cutoff \[P(\Delta r) = (\Delta r + \Delta r_0)^{ - \beta} \exp(\frac {\Delta r} {k})\]  (with $\beta = 1.75 \pm 0.15,~ \Delta r_0 = 1.5 $ km, and $k$ a cutoff value varying in different experiments)  and showed that individual truncated Levy trajectories coexist with population-based heterogeneity.\\
This heterogeneity was measured in terms of the radius of gyration $(r_g)$, which depicts the characteristic distance traveled by a user. It was shown \cite{barab} that the distribution of the radius of gyration can be approximated by a truncated power-law 
\[P(r_g) = {(r_g + r_g^0)}^{–\beta _{r}} \exp(–r_g/k)\] where $\beta r = 1.65 \pm 0.15, ~r ^0 _g = 5.8$ km and $k = 350$ km. In other words, most people usually travel in close vicinity to their home location, while a few frequently make long journeys.  
Radius of gyration as a key quantity in human mobility trajectories \cite{ellips} measures how far from center of the mass the trajectories are. It demonstrates the linear size occupied by each user's trajectory up up time $t$:
\[\overrightarrow{r_{g}^{a}}(t)= \displaystyle \sqrt{\frac{1}{n_{c}^{a}(t)} \sum_{i=1}^{n_{c}^{a}}(r_i^a - r_{cm}^{a})^2}  \]
where $\overrightarrow{r_{i}^{a}}$ represents the $i=1, . . . , n_{c}^{a}(t) $ positions recorded for user $a$ and $\overrightarrow{r_{cm}^a}= \displaystyle \frac{1}{n_{c}^{a}} \sum_{i=1}^{n_{c}^{a}}r_i^a$ is the center of the mass of the trajectory. Some studies have tried to model trajectories based on trajectories of individuals around their radius of gyration. 
Gonzalez et al \cite{barab} suggested using gyration radius as a characteristic travel distance for each individual. 
Xiao et al in \cite{ellips} with simplifying human mobility model with three sequential activities (commuting to workplace, going to do leisure activities and returning home), have been proved that daily moving area of individual is an ellipse, and they get an exact solution of the gyration radius. How ever, they have used some basic simplified assumptions which makes the model not to be used for all types of trajectories (it's not strong enough).\\ 
Xiao et al in \cite{divers} have been studying the trajectories of individuals in different categories (student/working group/not working group). Although the power law property of jump length distribution was observed in their study , but it also led to conclude that individual traveling process in general cannot be characterized by the Levy-flight or truncated Levy-flight.\\
\noindent As another property of human mobility, gravity models have been found in some studies  \cite{multiscale}, \cite{China}, \cite{uniModMob} which assumes the number of individuals $T_{ij} $ that move between location $i$ and $j$ per time unit is proportional to some power of the population of the source $(m_i)$ and destination $(n_j)$ location, and decays with the distance $t_{ij}$ between them as \[ T_{ij}= \frac{m_{i}^{\alpha} n_{j}^{\beta}}{f(r_{ij})} \]
($\alpha$ and $\beta$ are adjustable exponents and the deterrence function is chosen to fit the empirical data) Occasionally $T_{ij}$ is interpreted as the probability rate of individuals of traveling from $i$ to $j$, or an effective coupling between the two locations.\\
Having extracted spatial metrices, the correlation and relation between spatial measures is an interesting question in Human Mobility studies. Csaji et al in \cite{PCA} using data analysis techniques on a large data set of mobile phone users, illustrate that movement and location-related features are correlated with many other features and applying Principle Component Analysis (PCA) they show significant dimension reduction with limited loss of information is possible.
They have clustered users most common locations to home and office and estimated the position of frequent locations based on a probabilistic inference framework. Using the positions, they derived a fairly accurate distribution of the population, with a correlation of 0.92.  \\
\noindent As mentioned, \textbf{Pattern extraction} is one of the goals of studies on human mobility: to investigate if individuals movement follows any pattern and extract these patterns to define mobility models \cite{Barth} \cite{UnveilComplex} \cite{trajPatMin} \cite{PCA}.
These patterns could be spatial patterns \cite{UnveilComplex} \cite{PCA}, e.g. finding motifs in spatial network \cite{Barth},  
 spatio-temporal pattern \cite{UnveilComplex} or temporal patterns \cite{PCA}. Smoreda et al. in \cite{actpas} have been combined information on phone location and time of location to construct small movement oriented networks (places connected by trips), called mobility motifs. 
Surprisingly they found that the Paris phone traces and travel survey data reveal the same motifs, as do the Chicago survey data!

As a temporal pattern, daily mobility patterns have been illustrated in many \cite{persist} \cite{PCA} \cite{UnveilComplex}, this periodicity have been investigating in the distribution of different measures such as mean degree, mean clustering coefficient and the network adjacency correlation of a proximity network \cite{persist}, they could be found in hourly movement distribution \cite{UnveilComplex}, average calling dynamics \cite{PCA} or texting patterns \cite{clust}.

More precisely, ``trajectory pattern'' can be introduced \cite{trajPatMin} as a description of frequent behaviors which focuses on methods for extracting patterns  containing spatial and/or  temporal information.\\ 
Giannotti et al in \cite{UnveilComplex} have defined a set of patterns for trajectory data set which represent the common behavior of a (sub)-group of trajectories. They have defined T-pattern (a set of trajectories visiting the same -order of- regions in the same time interval), T-Cluster ( a set of trajectories grouped on the basis of their similarity according to a specified similarity function), T-folk (a spatio-temporal coincidence of a group of moving points in a specific time interval) and T-flow (the flow of a group of trajectories moving from one point to another).
They have also defined models of human mobility to describe their entire input data set. These models are mostly the collection or aggregation of patterns.  
Although a purpose of these approaches is to define mobility models by the set of distributions that fit some statistics extracted from the traces considered ,but it is important to notice these models do not propose a general mobility model that describes user's movements and their applicability outside the environment from which they have been derived is not clear \cite{Oppnet}


Some studies have been trying to answer if there is different mobility behavior among different group of users
\cite{China}, \cite{clust} , \cite{divers}. In China, \cite{China} women and children were generally found to travel shorter distances than men. Becker et al \cite{clust} have applied unsupervised clustering algorithm to CDR to investigate the groups of users where members of these groups share the same patterns of cell phone communication, in particular patterns of calling and texting intensity over time. Each group has a specific calling signature, which may be indicative of certain population types such as workers, commuters, and students. It would be an interesting question to answer that if these cluster of phone user have the same mobility behavior. 
In this part we reviewed the studies which have been trying to extract the patterns and models of Human Mobility. It was observed that a pattern is a representation of a local property that holds over a sub-group of mobility data and a model as a representation of a global property that holds over an entire data set, could be either a global aggregate or a collection of patterns \cite{UnveilComplex}.\\

In network sciences, mobility studies have recently become popular, especially using individuals trajectories based on mobile phone location data \cite{barab},\cite{limits}.The advantage of modeling the system as a graph is that we can say much about the behavior of the dynamical system without studying the actual dynamics \cite{tempNet}. It can be estimated how much one part of the network influences another and how well the network is optimized with respect to the dynamical system.
 In  Human Mobility studies, various networks could be defined. Transportation Network, road network, contact network, intercontact network are some examples.\\
A graph is a mathematical object consisting of a set of vertices, the units of the system, and a set of edges, the pairs of vertices that are interacting with each other \cite{tempNet}.
Within the scope of graph theory, mathematical measures such as centrality, connectedness, path length, diameter, degree and clique are playing key roles in network studies \cite{comNet}. \emph{Centrality} of a node, determines the relative importance of a node within the network.
It can summarize the ability of each node to broadcast and receive information. Centrality measure is known as one of mostly used parameters in network studies and thus, different types of centrality measures have been defined. 
According to \cite{centBook} there is no centrality index that fits all applications and the same network may be meaningfully analyzed with different centrality indices depending on the question to be answered. The authors in \cite{centBook} , have reviewed different centrality measures, such as degree centrality, family of betweenness centrality indices, closeness centrality indices, feedback centralities \cite{centBook}.

Using graph theory, there are a set of approaches that consider a particular class of networks which are embedded in the real space, i.e. networks whose nodes occupy a precise position, they are used to investigate the population flow, population study, etc. 
Base stations in cellular networks are instance of nodes for such networks. So is studying the voronoi diagram of the geographical positions
or railway station. In these networks density of population and flow of people can be studied. \\
In Human Mobility studies using networks, sometimes the topology of networks is changing by time and thus they are studied as a dynamic network such as contact networks in Human Mobility. 
The structure of contact patterns not only affects the spreading of disease, but this structure can also be exploited in controlling and preventing the spread  \cite{tempNet}.
Trying to investigate the dynamic evolution of human mobility, recently becoming an important part of researches in this area.
In dynamic studies of networks, the changes in network parameters and measures based on time variation are studied.
Thus, it could be said that dealing with time dimension to develop dynamic networks, is one of the main considerations. Temporal dimension of Mobility is one of the main aspect to develop models and  they lead to dynamic network analysis of human mobility.
The analysis of time-varying graphs is a relatively recent topic in network science, and is beginning to open up new avenues toward characterizing network dynamics \cite{evNet}.\\
As mentioned, understanding the temporal patterns of individual human interactions is essential in managing information spreading and in tracking social contagion. Human interactions (e.g., cell phone conversations and e-mails) leave electronic traces that allow the tracking of human interactions from the perspective of either static complex networks or human dynamics.
\cite{callPat}
This group of approaches focus on studying dynamic network of Human Mobility. This dynamic network could be the inter-contact network of people who are moving  to different places (the basic idea if \emph{Opportunistic Networks} whose goal is to enable communication in disconnected environments
\cite{Oppnet} ).The link in these networks illustrates a kind of relation between individuals. This relation  could be defined as the period of time during which two individuals are in mutual specified  range of distance or could be social contact among individuals (e.g.phone call). \cite{compNetAnHuM}, \cite{HMOpC}. 
For example in \cite{compNetAnHuM}, the structural properties of contacts are presented by a weighted contact graph, where the weights express how frequently and how long a pair of nodes is in contact.\\
In these types of networks the relation between social contact and  mobility patterns plays an important rule in human mobility studies. \\
These networks reflect the complex structure in people's movements: meeting strangers by chance, colleagues, friends and family by intention or familiar strangers because of similarity in their mobility patterns. 
The studies in these area have been tried to represent the complex resulting pattern of who meets whom, how often and for how long, in a compact and tractable way. This allows us to quantify structural properties beyond pairwise statistics such as inter-contact and contact time distributions.\\ 
As these networks are defined based on an interaction between individuals, they are not interesting to study individual trajectories.  they are good to study social behaviors or group activity  \cite{perPerio}.\\ 
There are many studies which have been investigating periodic behavior (patterns) of human mobility. Clause et al in \cite{perPerio} studied the temporal connectivity patterns in their data-set. Although their data set is poor, but they have shown that the persistence of proximity in their data set appears to consistently follow a heavy-tailed distribution. In order to investigate the periodicity of proximity (inter-contact) network, they have studied the adjacency of nodes in different time slots and measured the similarity between each two consecutive snapshots of network. Their spectral analysis  have been shown a strong daily periodic behavior. \\
Also  Jiang et al in \cite{callPat} have studied the temporal patterns of individual human interactions based on their calling data and the dynamics of calling patterns among cell phone users. They have investigated the communication patterns of cell phone users and after classifying them in different clusters, they have been studied different properties of each cluster of users. \\
In \cite{2scale} the authors have tried to investigate the human mobility dynamics of  scaled structured population
 by presenting a two scaled  human mobility model for a meta-population. The sub regions and regions are interlinked via intra-and inter regional transport network systems.
Defining a two-scale dynamic model, under various types of assumptions, different patterns of static behavior of the mobility process is studied and by the system of differential equations, the steady states of the mobility process were investigated.
 Their results reveal that the system has a natural tendency to quarantine itself without totally breaking a link in the transportation network system. They found a threshold point for the largest intra regional visiting time of residents of a given sub region that leads to either a total isolation of the residents from other sub regions within the region or a partial isolation of residents from some of  the sub regions within the region. 
Their work provides probabilistic and mathematical algorithmic tools develop different level nested type interaction rates as well as  network-centric dynamic equations. Also their analytic results demonstrated by simulation work with detailed figures exhibiting the type of steady-state population structures.\\
It was demonstrated that geographic distance plays an important role in the creation of new social connections: node degree and spatial distance can be combined in a gravitational attachment process that reproduces real traces. Instead, it was observed that links arising because of triadic closure, where users form new ties with friends of existing friends, and because of common focus, where connections arise among users visiting the same place, appear to be mainly driven by social factors. The authors in \cite{locBSN} have described a new model of network growth that combines spatial and social factors and reproduced the social and spatial properties observed in their traces. 
Becker et al in \cite{HMChar} patterns yields insights into a wide variety of important societal issues. For example, evaluating the impact of human travel on the environment depends on knowing how large populations move about in their daily lives. Similarly, understanding the spread of a disease
hinges on a clear picture of the ways that humans themselves move and interact. Other examples abound in
fields such as urban planning, where knowing how people come and go can help determine where to deploy infrastructure and how to reduce traffic congestion.\\
Centrality measure plays an important role in dynamic networks as well as static networks. Recently many studies have been trying to generalize this measure for dynamic networks ,\cite{evNet}, \cite{CenDy}, \cite{InfoFlow}, \cite{Tempo}, \cite{CharTemp}. 
The  \emph{average temporal path length} have been proposed in \cite{Tempo} and the characters of this temporal measure have been investigating and the new measure \emph{temporal reachability} is proposed based on average temporal path length in \cite{CharTemp}.
The concept of temporal closeness centrality is introduced in \cite{path} which is a generalization of closeness centrality.\\ 
Bagrow et al in \cite{mesos} show that individual mobility is dominated by small groups of frequently visited, dynamically
close locations, forming primary "habitats" capturing typical daily activity, along with subsidiary habitats representing additional travel. They try to investigate how similar are the habitats of users in close communication, and will this similarity be lower for pairs with less frequent interaction? they measure the similarity between the primary habitats of pairs of users interacting with one another by computing the relative number of locations the habitats have in common.
They define $P_{MFC}$, the probability that the next call placed by the user goes to that user's Most Frequent Contact and they study how $P_{MFC}$ depends on the properties of a use's mobility pattern. They showed that users who travel broadly, leading to complex mobility patterns and multiple habitats, tend to distribute their communication activity more uniformly over their contacts.

\subsection{Data collection techniques}
One of the basics in Human Mobility studies is data collection techniques as it indicates the accuracy of positioning system. 
Human mobility researchers have traditionally relied on expensive data collection methods, such as surveys and direct observation, to get a glimpse into the way people are moving. This high cost typically results in infrequent data collection or small sample sizes.
For example, a national census produces a wealth of information on where millions of people live and work, but it is carried out only once every ten years \cite{HMChar}. 
Brockmann et al. \cite{brock} used the data of bank notes to study human traveling behavior. 
Later on, many other studies used GPS (Global Positioning System) to track individuals or any moving objects \cite{uniModMob} \cite{barab} . GPS  provides  accurate measurements of both position and speed in outdoor locations (fine granularity of the location data), but signal quality is reduced or completely lost in indoor environments. Moreover, phone users tend to keep GPS turned off when not in use to avoid battery drain. When the GPS signal is available however, it tends to be a very good candidate for differentiating between dwelling and mobility \cite{wifi}. \\
Continuous scanning for WiFi APs has been used in context-aware computing to detect user mobility. This method is attractive because it can be performed on-line and in real-time, both desirable qualities for this class of applications \cite{wifi}.\\
%
In recent years, the emergence of information and communication technologies (ICTs), and substantial investments in wireless infrastructures have been led to extensive use of Call Data Records (CDR)in human mobility studies.
Although CDR may have some bias on Human mobility studies \cite{CDRbias} \cite{PCA}, but up to now, they have been providing the best data sets to study the human mobility  \cite{HMChar}.
Each CDR contains  the time a phone placed a voice call or received a text message, and the identity of the cellular antenna with which the phone was associated at that time. When joined with information about the locations and directions of those antennas, CDRs can serve as sporadic samples of the approximate locations of the phone's owner. CDRs are an attractive source of location information for three main reasons:\\ 1) They are collected for all active cellular phones, which number in the hundreds of millions of records.\\ 2) They are already being collected to help operate the networks, so that additional uses incur little marginal cost.\\ 3) They are continuously collected as each voice call and text message completes, thus enabling timely analysis. \\
On the other hands, CDRs have two significant limitations. One, they are sparse in time because they are generated only when a phone engages in a voice call or text message exchange. Two, they are coarse in space because they record location only at the granularity of a cellular antenna (with average error of 175 meter \cite{ctrack}). It is not obvious a priori whether CDRs provide enough information to characterize human mobility in any useful way \cite{HMChar}. \cite{GSM} The first problem could be solved if we modify   data collection system and track the user in fix time intervals.
Smoreda et al in \cite{actpas} describe two different data collection methods from cellular phone network: \textit{active} and \textit{passive} localization. Active localization provides a tool for
recording positioning data on a survey sample over a long period of time. Passive localization, on the other hand, is based on phone network data which are automatically recorded for technical or billing purposes (CDRs). \\ 
In \cite{wifi} the authors try to explore how visibility and signal strength of Global System for Mobile Communications (GSM) cell towers and WiFi beacons, which is already available on standard mobile handsets, can be used to generate mobility profiles. 
They have proposed the use of GSM data in order to avoid  privacy risk of "fine-grained" locations and other practical limitations of continuous GPS sampling such as reduced phone battery life, inconsistent coverage for typical users, and limited availability of integrated GPS in current mobiles phones.
Nowadays, the social network, temporal dynamics and mobile behavior of mobile phone users have often been analyzed independently from each other using mobile phone data sets \cite{trajPatMin}. 
Table 1 presents a summarized properties of different data collection methods.\\

\begin {table*}[ht]
\begin{center}

\begin{tabular}[!ht]{ |c|l|l| } 
\hline  
    \textcolor{blue!90}{\textbf{\textsf{Methods}}}& \textcolor{blue!90}{\textbf{\textsf{Advantages}}} &\textcolor{blue!90}{ \textbf{\textsf{Disadvantages}}} \\
  \hline    
\textcolor{blue!90}{{\textbf{\textsf{Survey \& direct }}}} & - Multi purposed use & - Expensive to collect data  \\
\textcolor{blue!90}{{\textbf{\textsf{ observations }}}} \cite{vtrack}   && - Not accurate\\ \hline
                                           & - Accuracy & - Low coverage area\\ 
 \textcolor{blue!90}{{\textbf{\textsf{Wi-Fi localization}}}} \cite{wifi} & - Energy usage $\sim $ 50\% GPS & - Providing access point is expensive\\ \hline
                                            & - Highly precise $\sim $ 5m error & - High battery (energy) usage  \\ 
 \textcolor{blue!90}{{\textbf{\textsf{GPS localization}}}} & - Can distinguish bettween & - Expensive\\ 
       \cite{actpas}, \cite{wifi} \cite{ctrack}  & - transportation modes   & - No (low quality) signal in indoor \\
                                             && invironment  \\ \hline
\textcolor{blue!90}{{\textbf{\textsf{Cellular network}}}} & & - Sparse in time  \\ 
\textcolor{blue!90}{{\textbf{\textsf{localization (passive)}}}} & - Automatically generated & - Needs more filtering\\
\textcolor{blue!90}{{\textsf{(Call Data Records)}}}\cite{HMChar}\cite{ctrack} &                             & - Less accuracy ($\sim $ 175 m error) \\ \hline
                                                    &- More accuracy than & - More costly than passive form \\
\textcolor{blue!90}{{\textbf{\textsf{Cellular network}}}} &  passive localization  & - Arise the issue of large database\\ 
\textcolor{blue!90}{{\textbf{\textsf{localization (active)}}}} & - Less expensive than & \\
                 \cite{actpas}                          &    previous methods    & \\ \hline

\end{tabular}
\caption {Comparative summary of different data collection techniques }
\end{center}
\end {table*}
\noindent It's important to notice that active localization method, by provoking cell localization of the mobile devise, solves the problem of time sparsity and provides a valuable resource of human trajectories. Considering the limitation of this type of data, which is spatial accuracy comparing other data types, it is important to chose a proper level of details to apply analysis on them.

\subsection{Human Mobility studies and Transportation}

Unlike other complex networks such as author citation, transportation networks are embedded in a metric space which raises a number of interesting questions such as:\\
How do the statistical properties of a network depend on scale size?
What are the differences between various transportation networks, and, more interestingly, what features do they share?
Do nodes play different roles in a network based on their connectivity? 
How can these roles be characterized? What can transportation networks tell us about the connectivity of spatially distributed communities? 
Transportation Networks  
govern many modern problems such as disease spread, congestion, urban sprawl, structure of cities \cite{Barth}. 
As Kuran et al. in \cite{Barth} present different representation of a transportation system. The simplest representation is obtained when the nodes represent the stations and links the physical connections.\\ 
Wang et al. in \cite{RoadPattern}  
found that the major usage of each road segment can be traced to its own - surprisingly few - driver sources. Based on this finding they propose a network of  road usage by defining a bipartite network framework,  demonstrating that in contrast to traditional approaches, which define road importance solely by topological measures, the role of a road segment depends on both: its betweennes and its degree in the road usage network.
 Moreover, the ability to pinpoint the few driver sources contributing to the major traffic flow allows us to create a strategy that achieves a significant reduction of the travel time across the entire road system, compared to a benchmark approach.\\


\noindent \textbf{Traffic Flow in Human Mobility}\\

One of the main purpose of studying human mobility in this concept is to investigate the population flow between different places \cite{Hunter2011} \cite{cong} with the aim of understanding and developing an optimal road network with efficient movement of traffic and minimal traffic congestion problems.\\
 In this part, main characteristics of traffic flows at the microscopic and macroscopic level are described . In a microscopic approach to traffic, each individual is examined separately. while at the macroscopic level we do not look at the individual as separate entities. The macroscopic level is more relevant to the dynamic description of traffic. Some macroscopic variables that translate the discrete nature of traffic into continuous variables are reviewed below:\\

\noindent \textbf{Density $(k)$} -
 Density is a typical variable from physics that was adopted by traffic science. Density $k$ reflects the number of vehicles per kilometer of road. For a measurement interval at a  certain point in time, such as $S_1$, $k$ can be calculated over a road section with $ \Delta X$ length as: \[ k=\displaystyle \frac{n}{\Delta X} \]
The index $n$ indicates the number of vehicles at $t_1$ on the location interval $\Delta X$. Total space 
of the $n$ vehicles can be set equal to $\Delta X$, thus:
\[ k= \displaystyle \frac{n}{\displaystyle \sum_i S_{i}}=\frac{1}{S}\]

	
\noindent \textbf{Flow rate $(q)$} - The flow rate $q$ can be compared to the discharge or the flux of a stream. The flow rate represents the number of vehicles that passes a certain cross-section per time unit. For a time interval $\Delta T$ at any location $x_2$, such as the measurement interval $S_2$ the flow rate is calculated as follows: 
\[q(x_2,t_2,S_2)= \displaystyle \frac{m}{\Delta T}\]
The index m represents the number of vehicles that passes location $x_2$ during $\Delta T$ . This time interval is the sum of the $m$ headways. Through the introduction of a mean headway $\overline{h}$ we find the following expression for the traffic flow rate:
\[ q=\displaystyle \frac{m}{\displaystyle \sum_i h_i} =\frac{1}{h} \]

 
\noindent \textbf{Mean Speed } - Mean speed $u$ is defined as the quotient of the flow rate and the density.  The mean speed is also a function of location, time and measurement interval. Note that the area of the measurement interval no longer appears in definition below:
\[u(x,t,S)= \displaystyle \frac{q(x,t,S)}{k(x,t,S)}\] 
In another word, it is the fraction of \emph{Total~distance~coveredby~vehicle~s~in~ S} to the \emph{Total~ time ~spent~by~vehicle~s~in~S}\\
The definition of the mean speed is also called the fundamental relation of traffic flow theory: 
\[q=k\cdot u\] This relation irrevocably links flow rate, density and mean speed. Knowing two of these variables immediately leads to the remaining third variable.  
This equation incorporates the interdependence of traffic flow, traffic density  and speed. When two of the three variables are known, the third variable can easily be obtained. If traffic
count data are available, traffic flows can be assumed as given, which leaves us to calculate either
traffic density or speed to complete the formula and use either as input for the appropriate
queuing model.


The macroscopic traffic variables can be calculated for every location, at any point in time and for every measurement interval. In practice we mostly use traffic detectors that measure the macroscopic variables $u$ and $q$ across a certain time interval. 
 The discrete nature of traffic requires time intervals of at least half a  minute if we want to achieve meaningful information. When the time intervals exceed a duration of five minutes, some dynamic characteristics are lost. \\
Traffic flow in microscopic scale, can be investigated by queuing theory. Queuing and Traffic Flow is the study of traffic flow behaviors surrounding queuing events.\\
Beside these notation of traffic flow, \emph{Centrality} measure plays an important role in these studies. Centrality of a node determines the relative importance of a node within the graph. Different types of centrality measures have introduced in recent studies which we mention three classic centrality measures below. Having graph $G = (V,E)$ with $V$ a set of nodes of a graph  and $E$ the set of edges in $G$ and $a_{ij}=a_{ji}$ showing the link between node $i$ and node $j$,\\
1) Degree centrality-  Degree centrality is defined as the number of links incident upon a node.  The idea is that a node with more edges is more important : \[C_{Degree}(N_i) =  \sum_{j=1}^n a_{ij} \]\\
2)Closeness centrality- measures the importance of a node by its geodesic distance to other nodes. The idea is the closer a node is to other nodes, the important the node is.
Closeness can be regarded as a measure of how long it will take information to spread from a given vertex to others in the network . Closeness centrality focuses on the extensivity of influence over the entire network. \ \[ C_{Closeness}(N_i)=  \displaystyle \frac{1}{ \displaystyle \sum _{i=1} ^ Nd(N_i,N_j)}\]Where $d(N_i, N_j )$ is a geodesic distance between $N_i$, and $N_J$.\\
3)Betweenness centrality-  measures the importance of a node by its proportion of paths between other nodes. The idea is that a node that plays the roles of connecting more other nodes is more important. It explains how a node can control the other nodes which have no direct connectivity between them.
\[ C_{Betweenees} (N_i)= \sum_{j<k} \frac{g_{jk}(N_i)}{g_{jk}} \] Where $g_{jk}$ is the number of
 geodesic paths between two nodes $N_j$ and $N_k$, and $g_{jk}(Ni)$ is the number of geodesic 
between the $N_j$ and $N_k$ that contain node $N_i$. \\

\begin{figure*}[ht]
 
  \centering
    \includegraphics[scale=0.45]{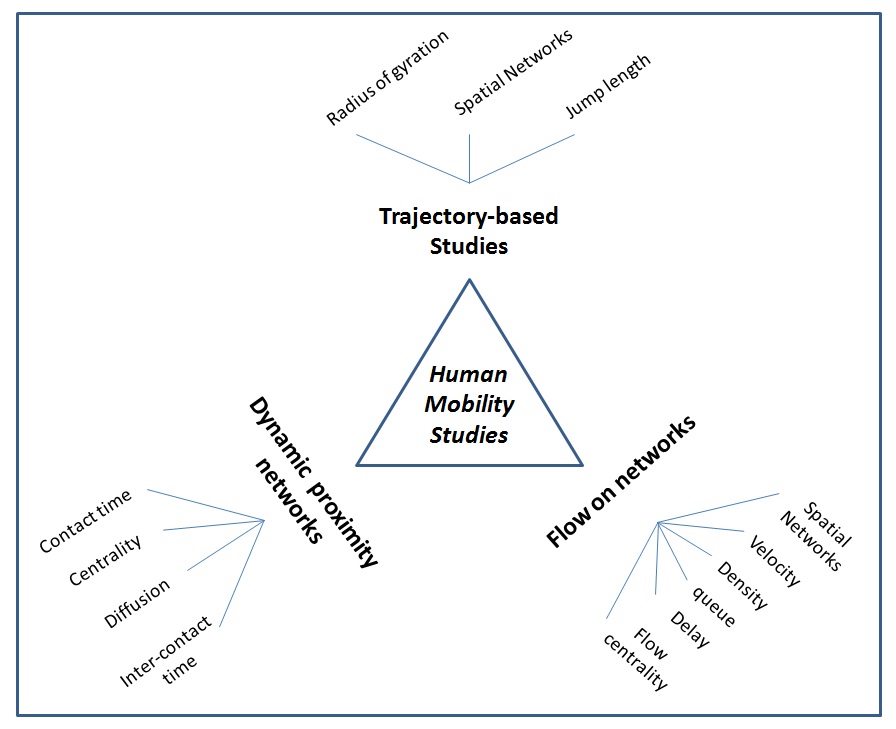}
    \caption{The three main baselines in Human Mobility studies}
\end{figure*}

\noindent The relation between congestion and centrality in traffic flow was studied by Petter Holme in \cite{cong}. His work investigates the relation between centrality assessed from the static network structure measured in simulations of some simple traffic flow models. 
He studied how the speed of the traffic flow is affected by the network structure (by tuning model parameters) and found that the relationship between the betweenness centrality and congestion in simple particle hopping models for traffic flow. 
Altshuler et al. in \cite{AugCent} have studied the relationship between centrality of a node and its expected traffic flow in transportation networks. In their experiment using the dataset that covers the Israeli transportation network, they show the correlation between the traffic flow of nodes and their Betweenness centrality and they also show that when some additional known properties of the links (specifically, time to travel through links) are taken into account, this correlation can be  significantly increased 
which could be used to generate highly accurate approximations of the traffic flow in the network.\\
Authors in \cite{clust} have tried to produce accurate models of how large populations move within different metropolitan areas with the goal of generating sequences of locations and associated times that capture how individuals move between important places in their lives, such as home and work. They have tried to find a model for mobility data in metropolitan areas in order to help address important societal issues such as the environmental impact of home-to-work commutes. They also tried to take into account how different metropolitan areas exhibit distinct mobility patterns due to differences in geographic distributions of homes and jobs, transportation infrastructures, and other factors.\\
Network centrality of metro systems in different countries have been studied applying the notion of betweenness centrality to 28 worldwide metro systems \cite{Metrocent}. The share of betweenness was found to decrease in a power law with size (with exponent 1 for the average node), but the share of most central nodes decreases much slower than least central nodes. 
 The betweenness of individual stations as a node can be useful to locate stations where passengers can be redistributed to relieve pressure from overcrowded stations.\\
\emph{Edge Centrality} is another important metric in studying flow through the network \cite{CN}. There is also a group of studies which aim to use Expectation Maximization algorithm for estimating historical link travel time distributions across an arterial road network \cite{pathNtravel} \cite{Hunter2011}.
Sun et al. \cite{SunWSMYS10} in a novel approach proposed to explore the space-time structure of urban dynamics, based on the original data collected by cellular networks in a southern city of China, recording population distribution by dividing the city into thousands of pixels. By applying principal component analysis, the intrinsic dimensionality is revealed. The structure of all the pixel population variations could be well captured by a small set of eigen pixel population variations. According to the classification of eigen pixel population variations, each pixel population variation can be decomposed into three constitutions: deterministic trends, short-lived spikes, and noise. Moreover, the most significant eigen pixel population variations are utilized in the applications of forecasting and anomaly detection.

In these parts we presented different studies and findings about Human Mobility. According to reviewed findings, we distinguish three different baselines in
 Human Mobility studies. The first categories belongs to trajectory-base studies, which is mainly on analyzing trajectories of individuals. In these set of approaches, statistical measures such as jump length and radius of gyration  are analyzed in order to find patterns and models for individuals movement. 
The second category is studying dynamic proximity networks of people in mobile ad hoc networks. 
In such a network 
, finding a route between two disconnected devices implies uncovering habits in human movements and patterns in their connectivity (frequencies of
 meetings, average duration of a contact, etc.), and exploiting them to predict future encounters \cite{Oppnet}. The third category of Human Mobility 
 studies related to investigating  flow on networks such as  road network, transportation network, infrastructure networks and etc. 
Figure 1 illustrates three different baselines that we have distinguished for Human Mobility studies.

\subsection{Tools and Applications}
In mobility studies, if we want to analyze the real time data, computation and data storage will be a concern. 
One of the main concerns  is about data storage issues: how to store trajectory data in order to optimize different operations on the mobility data. 
Giannotti et al. in \cite{UnveilComplex} have presented a  mobility data mining tool, a \textbf{query language} centered onto the concept of trajectory.
Their  Query language, has been designed to store, create and present trajectory data, patterns and models. 
\textbf{Graph Database} is another approach to store networked data and perform efficient computation on such type of data set. As an example of using graph database in Human Mobility studies, we can mention  Daytsher et al \cite{neo4j} who  have tried to extract the life pattern edges synopsize the location history of people, and accordingly, connect individuals to places they frequently visit, based on Graph Data base using Neo4j as the underline database-management system. \\
In order to have a near real time system  which is able to predict the future flow/situation based on the historical data of trajectories, one solution is to use distributed algorithm which uses a streaming of real data to perform the calculation and make prediction.
Hunter et al. \cite{Hunter2011}, have presented an experience scaling up the Mobile Millennium traffic information algorithm in the cloud. Their study affirmed the value of in-memory computation for iterative algorithms, but also highlighted three challenges that have been less studied in the systems literature: efficient memory utilization, broadcast of large parameter vectors, and integration with of-cloud storage systems. These factors were found crucial for performance.\\
Regarding traffic engineering,there is a set of tools which use mobility data to model traffics and mainly used in urban planning and  traffic engineering. They try to simulate traffic models based on vehicles movement. SUMO (Simulation of Urban MObility) \cite{sumo}, MobisSim \cite{mobisim}
, VanetMobiSim 
, CANU Mobility Simulation Environment 
and 
ns (network simulator) 
are some examples of such tools.\\

\begin {table*}[ht]
\begin{center}

\begin{tabular}[!ht]{ |c|c|c|c|c|c|c|c|c|c|c|c|c| } 
\hline  
 & \multicolumn{3}{|c|}{\textcolor{blue!90}{\textsf{\textbf{Scale}}}}  & \multicolumn{3}{|c|}{\textcolor{blue!90}{\textbf{\textsf{Study Baseline}}}} &\multicolumn{3}{|c|}{\textcolor{blue!90} {\textbf{\textsf{Metrics}}}}& 
 \multicolumn{3}{|c|}{\textcolor{blue!90}{\textbf{\textsf{Data type}}}}
 \\  \hline

&\begin{sideways} Country scale \end{sideways}&\begin{sideways} Regional scale \end{sideways}&\begin{sideways} City scale \end{sideways}&\begin{sideways} Trace-based studies \end{sideways}&\begin{sideways} Dynamic proximity networks \end{sideways}
&\begin{sideways} Flow on network  \end{sideways} &\begin{sideways} Centrality measure \end{sideways} &\begin{sideways} Jump length distribution \end{sideways} &\begin{sideways} Radius of gyration \end{sideways} &\begin{sideways} GPS \end{sideways} &\begin{sideways} GSM data \end{sideways} &\begin{sideways} Wifi \end{sideways}\\
  \hline
   & \cite{PCA} & \cite{China} & \cite{UnveilComplex} & \cite{UnveilComplex} & \cite{callPat}&\cite{cong} &\cite{pathNtravel} &\cite{barab} &\cite{ellips}&\cite{Hunter2011}  & \cite{actpas}& \cite{vtrack} \\

   &\cite{actpas} &\cite{barab}& \cite{Hunter2011} &\cite{PCA} &\cite{perPerio} &\cite{Hunter2011}&\cite{Hunter2011}&\cite{divers}&\cite{barab}& \cite{UnveilComplex} &\cite{ctrack}&\\

\textbf{Related}     & \cite{vespignani}&\cite{uniModMob}& \cite{HMChar}& \cite{actpas} & &\cite{cong}& \cite{cong} &\cite{brock}&\cite{PCA}& \cite{trajPatMin}  &\cite{PCA}&\\

\textbf{studies}  &&&\cite{RoadPattern} & \cite{ctrack}& &&\cite{Metrocent}&\cite{ellips}&&  \cite{divers} &\cite{uni}&\\
           &&&\cite{trajPatMin} & \cite{vtrack}& &&&&  &  &\cite{barab}&\\
           &&& & \cite{Hunter2011}& &&&&&                 &&\\
           &&& & \cite{ellips}& &&&&&                 &&\\
           &&& & \cite{uniModMob}& &&&&&                 &&\\
           &&& & \cite{China}& &&&&&             &&\\
\hline

\end{tabular}
\caption {Summary table }
\end{center}
\end {table*}

\subsection{Discussion}
 In this part we try to point some current challenges in human mobility studies and the approaches that can be taken to solve the problems in each case, are discussed.\\

\textbf{Data pre-processing} is step zero in almost any mobility studies. This step extremely depends on the data type and the purpose of the analysis.   
Active data localization \cite{actpas} provides us with the frequent record of the mobile users and higher quality of raw data that are more powerful to be used in human mobility modeling/analysis \cite{actpas}, but the granularity problem of the data (since the locations are recorded only at the granularity of the cellular antenna) still remains. The previous studies that have been using the human mobility trajectories in traffic studies, have been used GPS data \cite{Hunter2011} \cite{UnveilComplex} \cite{trajPatMin} \cite{divers} which is much more accurate and suitable to be used for map matching. 
The data collection techniques also could help to identify special type of users (like GPS data of drivers), while in active localization method there is no information about transportation mode of the user which have to be extracted from data. All these reasons, encounter us with a new version of mobility data which arises many challenges.  
Smoreda et al. in \cite{actpas} propose a map-matching algorithm to identify different modes of transportation. The algorithm matches mobile phone location data and vector data (e.g. railways, roads, airports) using the two main steps described below.
(1) The local approach,which entails matching the recorded mobile phone locations to the correct vector data based on two criteria: Euclidean distance and speed between two consecutive recorded locations. (2) The global approach: this step consists in identifying the route traveled and the mode of transportation. The assumption with the highest number of points is chosen. Due to the fact that there is no uncertainty at this level, a confidence measure is defined in this approach.
  
 One possible approach would be using pattern matching algorithm to project the trajectory data points to the network which has the minimum error rate. The authors in \cite{Hunter2011} have been using Expectation Maximization algorithm to do this projection.\\
 Thiagarajan et al. in \cite{ctrack} have used the additional data of accelerometers (to detect movement) and magnetic compasses (to detect turns) with active cellular
  network data and using a two-pass Hidden Markov Model \cite{vtrack} they have achieved a good accuracy of map-matching result. considering that they have collected the
   data of car drivers, the question of accuracy of this method applied on a large actively collected cellular network data still remains. The combination of this method and two level
    matching algorithm presented by Smoreda \cite{actpas} could help us to achieve a suitable result for map-matching problem.    
 The other approach is to use a combination of different networks: the idea of \emph{Multiplex network} presents a multiplex network \cite{multiplex} that each network has to
be considered as being a part of a larger system in which a set of interdependent networks with different structure and function coexist, interact and coevolve. The structural properties
 of each of these networks and their evolution can depend in a non-trivial way on that of other graphs to which they are interconnected. \\
A current challenge in Human Mobility studies is  mapping individual trajectories to a multiplex network of on-ground and underground transportation networks.
An orientation of this research could be studying human mobility using multi-layer  transportation network and also investigating the interdependency between
 different layers in transportation networks. Investigating degree and  coefficient relationships for different layers of layered network is an instance that Wou et al.
 in  \cite{layered} have discussed. 
 Other related networks to the transportation network, could be added and investigated in multiplex networks (e.g. infrastructural networks, social networks). 

\textbf{Mobility Distribution} 
 Barabasi et al in \cite{barab} showed that jump length of human mobility follows power law, 
 Blondel's study \cite{PCA} in Portugal showed  , travel distances exhibit a log-normal distribution for $d < 150$ km.
Yan et al. in \cite{divers} studied specifically travel distance of different cluster of individuals and they observed that for the aggregated population, the jump length distribution follows a power law with an exponential cutoff. While they didn't find such property analyzing different user clusters (students, employees, retirees). Yan et al. \cite{divers} also suggest that the form (power law, exponential) of deterrence function in the gravity law of human travel, may be sensitive to the mode of transportation under consideration. Montjoye et al. also in \cite{uni} mention the parameter that should be consider in generalizing the result of studies on mobility data to larger population.\\
 It can be investigated if different categories of mobility, follow the same statistics rules or they could be distinguished with different distributions. We also aim to study and validate the existing findings and models on our new interesting data set. \\

\textbf{Behavioral dimension in Human Mobility}  Csaji et al. in \cite{PCA} 
extracted a variety of spatial features of human mobility from their trajectory data and used these features to cluster different types of mobility behavior. 
Zignani et al in \cite{beh} have tried to develop a model to infer the probability distributions of all the features of human behavior by analyzing human mobility traces.

Marcel Hunecke in  figure 2 \cite{adhome}, have defined two different set of factors of influence on individual mobility behavior:
\emph{Personal factors} and \emph{External factors}. Among these factors, two types of personal factors are mentioned as relevant factors for individual mobility: socio-demographic characteristics and attitudinal factors. Socio-demographic aspects include factors, which determine individual options and needs for mobility activities. 
Attitudinal factors include values, norms and attitudes (e.g. symbolical estimations of transport modes), which affect preferences and habits for specific activities, destinations, routes and means of transport \cite{adhome}.\\
An orientation of the research could be trying to define and extract behavioral feature of individuals from individual mobility traces (such as popular places, the media and content they have accessed via internet, working pattern and etc.)  
If we could demonstrate behavioral features extracted from mobility data can be used to categorize Mobility behaviors , it opens an additional door in studying human mobility aspects:
 we could claim that  another additional "Behavioral" dimension: can be added to dimensions of Human Mobility presented by Karamshuk \cite{Oppnet}.\\ 
Obviously in these types of study, privacy of mobile users is an important concern \cite{priv}, and should be clearly specified.\\
Feature extraction from mobility data, is a challenging issue which requires wide range of data analysis. Various data types discussed in section
 2.1 is another reason of this challenges. For instance, transportation mode of the user is an important feature which helps with later studies on user behavior
  studies. Many data sets used in Human Mobility studies, have collected by a distinct group of users.  Cellular network data which collects data of a huge anonymous 
  group of users,  brings us the hidden valuable features and provide us mobility data with weak accuracy in spatial 
   positioning. Extracting transportation mode from this data is an important challenge which we aim to investigate which is in data pre-processing
    part.\\
\begin{figure*}[ht]
 
  \centering
    \includegraphics[scale=0.45]{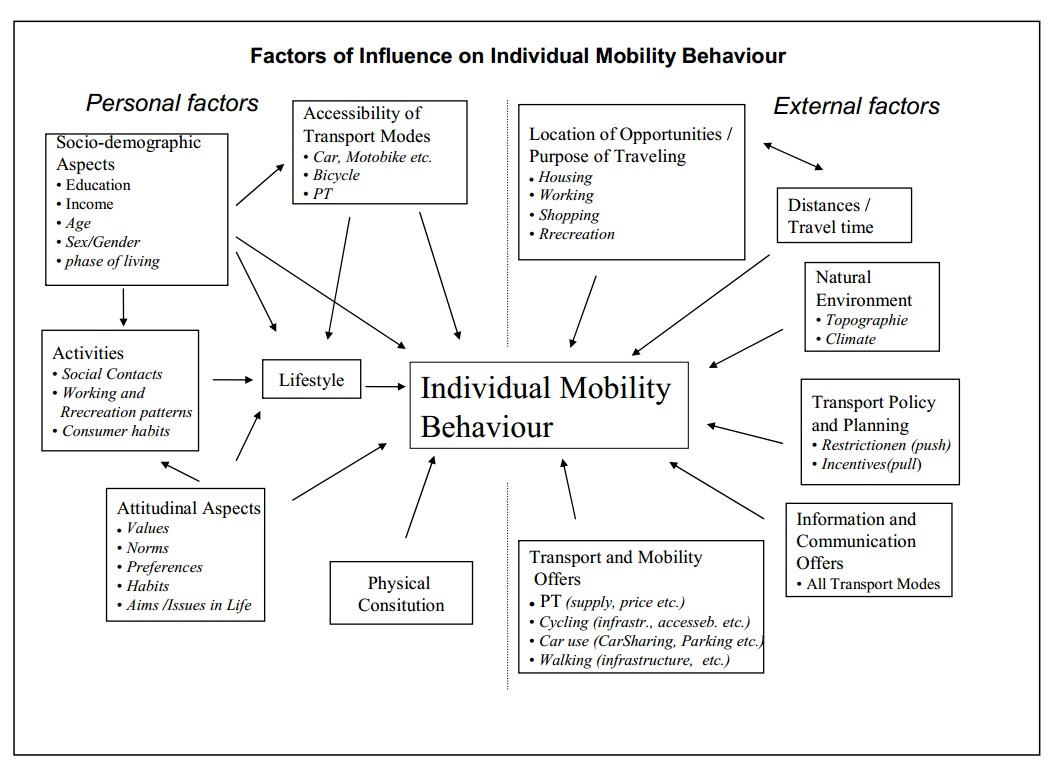}
    \caption{factors of influence on individual mobility behavior, source: \cite{adhome}}
\end{figure*}

\textbf{Graph data base for Human Mobility} As mentioned before, storage system of trajectory data and applying analysis on it, is a concern in Human
 Mobility studies. Graph Database was introduced mainly to represent and work with networked data where there are some relationships defined for links of the
  network. It has been shown that some computation and evaluation which are graph related measures (such as shortest path or pattern finding models) are
   faster and more practical than result of relational database. This approach can be implemented on trajectory data and the efficiency of this approach could
    be evaluated for comparing with other algorithms.


\section{Perspectives and Conclusion}
In this survey we presented a classified summary of Human Mobility studies based on the context of the study. We mainly distinguished three different baselines in Human Mobility studies . It was discussed that new type of mobility data bring more opportunities for mobility studies while they also have some challenges regarding the granularity of data which affects the evaluation when we are working on human mobility in city scale. These issues bring the new pre-processing issues to the context of Human Mobility as well as novel approaches in data analysis and feature extractions.
Later on, we aim to proceed in applying new type of data (active localization cellular network data) on mobility studies. Detecting transportation mode is a new feature that is concerned in these new type of collected data. We try to apply presented model and derive a new combined algorithm to derive a new solution for this issue.  The behavioral aspect of Human Mobility was proposed as a new dimension of Human Mobility and we could extend the presented models with the new dimension if we can illustrate the correlation between behavioral aspects and mobility patterns of individuals. 

\bibliographystyle{plain}
\bibliography{Ref} 

\begin{thebibliography}{10}

\bibitem{multiscale}
Duygu~Balcan ~, Vittoria Colizza et~al.
\newblock Multiscale mobility networks and the spatial spreading of infectious
  diseases.
\newblock {\em PNAS}, 2009.

\bibitem{mesos}
James~P.Bagrow at~al.
\newblock Mesoscopic structure and social aspects of human mobility.
\newblock July 2012.
\newblock DOI: 10.1371/journal.pone.0037676.

\bibitem{Barth}
Marc Barth.
\newblock {Spatial Networks}.
\newblock pages 1--86.

\bibitem{valImp}
Rahul~C. Basole.
\newblock The value and impact of mobile information and communication
  technologies.
\newblock {\em Proceedings of the IFAC Symposium, Atlanta, GA.}, 2004.

\bibitem{sumo}
M.~Behrisch, L.~Bieker, J.~Erdmann, and D.~Krajzewicz.
\newblock Sumo - simulation of urban mobility: An overview.
\newblock In {\em SIMUL 2011, The Third International Conference on Advances in
  System Simulation}, pages 63--68, Barcelona, Spain, October 2011.

\bibitem{mAspects}
Masao~Kakihara Carsten~Sorensen.
\newblock Expanding the 'mobility' concept.
\newblock {\em SlGGROUP Bulletin}, 22(3), 2001.

\bibitem{limits}
Nicholas Blumm Albert-Laszlo~Barabási Chaoming~Song, Zehui~Qu.
\newblock Limits of predictability in human mobility.
\newblock {\em Science}, 327.

\bibitem{CN}
Reuven Cohen and Shlomo Havlin.
\newblock {\em {C}OMPLEX {N}ETWORKS Structure, Robustness and Function}.
\newblock {C}AMBRIDGE University Press, 2010.

\bibitem{uni}
Yves-Alexandre de~Montjoye~et al.
\newblock Unique in the crowd: The privacy bounds of human mobility.
\newblock {\em Scientific REPORTS}, 1376(3), March 2013.
\newblock DOI:10.1038/srep01376.

\bibitem{Metrocent}
Sybil Derrible.
\newblock Network centrality of metro systems.
\newblock {\em PlosONE}, 7(7).

\bibitem{UnveilComplex}
Fosca~Giannotti Dino~Pedreschi, Mirco~Nanni.
\newblock Unveiling the complexity of human mobility by querying and mining
  massive trajectory data.
\newblock {\em The international Journal on Very Large Data Bases}, July 2011.
\newblock Available online in
  \url{http://didawiki.cli.di.unipi.it/lib/exe/fetch.php/dm/vldbj_kddlab_matlas.pdf}.

\bibitem{centBook}
Leon Peeters-Stefan Richter Dagmar Tenfelde-Podehl Dirk~Koschutzki, Katharina
  Anna~Lehmann and Oliver Zlotowski.
\newblock {\em Chapter 3:Centrality Indices}.
\newblock Springer-Verlag Berlin Heidelberg.

\bibitem{persist}
Aaron~Clauset et~al.
\newblock Persistence and periodicity in a dynamic proximity network.
\newblock {\em Proceeding of DIMACS Workshop on Computational Methods for
  dynamic Interaction Network}.

\bibitem{perPerio}
Aaron~Clauset et~al.
\newblock Persistance and peridicity in a dynamic proximity network.
\newblock {\em DIMACS workshop on Computational Methods for Dynamic Interaction
  Neworks}, 2007.

\bibitem{tale}
Anastasion~Noulas et~al.
\newblock A tale of many cities: universal patterns in human urban mobility.
\newblock {\em Physics}.

\bibitem{vtrack}
Arvind~Thiagarajan et~al.
\newblock Vtrack: Accurate, energy-aware road traffic delay estimation using
  mobile phones.
\newblock {\em SenSys 09}, November 2009.

\bibitem{ctrack}
Arvind~Thiagarajan et~al.
\newblock Accurate, low-energy trajectory mapping for mobile devices.
\newblock {\em NSDI'11 Proceedings of the 8th USENIX conference on Networked
  systems design and implementation}, 2011.

\bibitem{PCA}
Balazs~Cs.Csaji et~al.
\newblock Exploring the mobility of mobile phone users.
\newblock {\em Statistical Mechanics and its Applications}, 392.

\bibitem{brock}
D.~Brockmann et~al.
\newblock The scaling laws of human travel.
\newblock {\em Nature}, 439, January 2006.

\bibitem{GSM}
Daniel~Schulz et~al.
\newblock Human mobility from gsm data - a valid alternative to gps?

\bibitem{neo4j}
Doytsher et~al.
\newblock Managing socio-spatial data as large graphs.
\newblock {\em WWW 2012 Companion}, April 2012.

\bibitem{vespignani}
Duygu~Balcana et~al.
\newblock Multiscale mobility networks and the spatial spreading of infectious
  diseases.
\newblock {\em PNAS}, 106(51), 2009.

\bibitem{hist}
Fosca~Giannotti et~al.
\newblock A complexity science perspective on human mobility, 2012.

\bibitem{CDRbias}
Gyan~Ranjan et~al.
\newblock Are call detail records biased for sampling human mobility?
\newblock {\em ACM SIGMOBILE Mobile Computing and Communications Review}.

\bibitem{Sptmp}
Hang-Hyun~Jo et~al.
\newblock Spatiotemporal correlation of handset-based service usages.
\newblock {\em EPJ Data Science}.

\bibitem{HMmetro}
Isaacman et~al.
\newblock Human mobility modeling at metropolitan scales.
\newblock {\em MobiSys}, 2012.

\bibitem{RoI}
Jing~Yuan et~al.
\newblock Discovering region of diffferent functions in a city using human
  mobility and pois.
\newblock {\em ACM KDD}, pages 186--194, 2012.

\bibitem{beh}
Matteo~Zignani et~al.
\newblock Extracting human mobility and social behavior from location-aware
  traces.
\newblock {\em Wireless Communications and Mobile Computing}, 13.

\bibitem{locBSN}
Miltiadis~Allamanis et~al.
\newblock Evolution of a location-based online social network: Analysis and
  models.
\newblock {\em IMC '12 Proceedings of the 2012 ACM conference on Internet
  measurement conference}, pages 145--158, 2012.

\bibitem{wifi}
Min Y.~Mun et~al.
\newblock Parsimonious mobility classification using gsm and wifi traces.
\newblock {\em HotEmNets}, 2008.

\bibitem{empS}
Ming~Zhao et~al.
\newblock Empirical study on human mobility for mobile wireless networks.
\newblock {\em Military Communications Conference, 2008. MILCOM 2008. IEEE},
  November 2008.

\bibitem{evNet}
Peter~Grindrod et~al.
\newblock Communicability across evolving networks.
\newblock {\em Physical Review}, 83:046120, 2011.

\bibitem{cong}
Petter~Holme et~al.
\newblock Congestion and centrality in traffic flow on complex network.
\newblock {\em Advances in Complex Systems}, pages 163--176, December 2003.
\newblock DOI: 10.1142/S0219525903000803.

\bibitem{RoadPattern}
P.Wang et~al.
\newblock Understanding road usage patterns in urban areas.
\newblock {\em Sci}, 2(101), December 2012.
\newblock doi:10.1038/srep01001.

\bibitem{clust}
Rechard~A.Becker et~al.
\newblock Clustering anonymized mobile call detail records to find usage
  groups.
\newblock 2011.

\bibitem{HMChar}
Richard A.~Becker et~al.
\newblock Human mobility characterization from cellular network data.
\newblock {\em Communications of the ACM}, 56(1):74--82, January 2013.

\bibitem{layered}
Sheng-Rong~Zou et~al.
\newblock Topological relation if layered complex networks.
\newblock {\em Physics Letters A}, 374:4406--4410, 2010.

\bibitem{China}
Tini~Garske et~al.
\newblock Travel patterns in china.
\newblock {\em PLoS ONE}, 6(2).

\bibitem{multiplex}
V.~Nicosia et~al.
\newblock Growing multiplex networks.
\newblock {\em Submitted in Physics and Society}, February 2013.

\bibitem{ellips}
Xiao-Yang et~al.
\newblock Exact solution of gyration radius of individual's trajectory for a
  simplified human mobility model.

\bibitem{divers}
Xiao-Yong~Yan et~al.
\newblock Diversity of individual mobility patterns.
\newblock November 2012.

\bibitem{AugCent}
Yaniv~Altshuler et~al.
\newblock Augmented betweenness centrality for mobility prediction in
  transportation networks.
\newblock {\em International Workshop on Finding Patterns of Human Behaviors in
  NEtworks and MObility Data ,NEMO'11}.

\bibitem{actpas}
Zbigniew~Smoreda et~al.
\newblock Spatiotemporal data from mobile phones for personal mobility
  assessment.
\newblock {\em International conference on transport survey Methods: Scoping
  the Future while Staying on Track}, 2013.

\bibitem{callPat}
Zhi-Qiang~Jiang et~al.
\newblock Calling patterns in human communication dynamics.
\newblock {\em PNAS}, 110(5):1600--1605, January 2013.
\newblock Available online in
  \url{http://www.pnas.org/content/110/5/1600.full.pdf}.

\bibitem{optReso}
Michele Coscia Salvatore~Rinzivillo Fosca~Giannotti, Dino~Pedreschi.
\newblock Optimal spatial resolution for the analysis of human mobility.
\newblock {\em IEEE/ACM International Conference on Advances in Social Network
  Analysis and Mining}, 2012.

\bibitem{priv}
Fosca Giannotti and Dino Pedreschi, editors.
\newblock {\em Mobility, Data Mining and Privacy - Geographic Knowledge
  Discovery}.
\newblock Springer, 2008.

\bibitem{2scale}
Divine~Wanduku G.S.~Ladde.
\newblock A two-scale network dynamic model for human mobility process.
\newblock {\em Mathematical Biosciences}, 229, December 2010.

\bibitem{NonDy}
Dirk~Brockmann Heinz Georg~Schuster.
\newblock {\em Reviews of Nonlinear Dynamics and Complexity,}.
\newblock John Wiley \& Sons, July 2010.

\bibitem{adhome}
ADD HOME.
\newblock Mobility management and housing project.
\newblock 2006-2009.
\newblock
  \url{http://add-home.eu/docs/Factors_Influence_Mobility_Behaviour.pdf}.

\bibitem{pathNtravel}
Timothy Hunter.
\newblock Path and travel time inference from gps probe vehicle data.
\newblock {\em NIPS Analysing Networks and Learning with Graphs}.

\bibitem{Hunter2011}
Timothy Hunter, Teodor Moldovan, Matei Zaharia, Samy Merzgui, Justin Ma,
  Michael~J. Franklin, Pieter Abbeel, and Alexandre~M. Bayen.
\newblock {Scaling the mobile millennium system in the cloud}.
\newblock {\em Proceedings of the 2nd ACM Symposium on Cloud Computing - SOCC
  '11}, pages 1--8, 2011.

\bibitem{Oppnet}
Opics In.
\newblock {Human Mobility Models for Opportunistic Networks}.
\newblock (December):157--165, 2011.

\bibitem{tempNet}
Petter~Holme Jari~Saramaki.
\newblock Temporal networks.
\newblock {\em Physics Reports}, 519:97--125, October 2012.

\bibitem{path}
Raj~Kumar Jari~Saramaki.
\newblock Path lengths, correlations, and centrality in temporal networks.
\newblock {\em Phys. Rev. E 84}, 2011.
\newblock Available online in \url{http://arxiv.org/abs/1101.5913}.

\bibitem{CharTemp}
Cecilia Mascolo Vito~Latora John~Tang, Mirco~Musolesi.
\newblock Characterising temporal distance and reachability in mobile and
  online social networks.
\newblock {\em ACM SIGCOMM Computer Communication Review}.

\bibitem{Tempo}
Cecilia Mascolo Vito~Latora John~Tang, Mirco~Musolesi.
\newblock Temporal distance metrics for social network analysis.
\newblock {\em in Proceedings of the 2nd ACM SIGCOMM Workshop in online Social
  Networks (WOSN09)}.

\bibitem{InfoFlow}
Cecilia Mascolo Vito Latora Vincenzo~Nicosia. John~Tang, Mirco~Musolesi.
\newblock Analysing information flows and key mediators through temporal
  centrality metrics.
\newblock {\em In Proceedings of 3rd Workshop on Social Network Systems (SNS)},
  April 2010.
\newblock Available online in:
  \url{http://www.cs.bham.ac.uk/~musolesm/papers/sns10.pdf}.

\bibitem{HMOpC}
Pan~Hui Jon~Crowcrof.
\newblock Human mobility models and opportunistic communication system design.
\newblock {\em Mathematical, Physical and Engineering Sciences},
  336(1872):2005--2016, 2008.

\bibitem{CenDy}
Jeon Hyung~Kang Kristina~Lerman, Rumi~Ghosh.
\newblock Centrality metric for dynamic networks.
\newblock 2010.

\bibitem{uniModMob}
Amos Maritan Albert-Laszlo~Barabasi Marta C.~Gonzalez, Filippo~Simini.
\newblock A universal model for mobility and migration patterns.
\newblock {\em Nsture}, 484:96--100, 2012.

\bibitem{barab}
Albert-Laszlo~Barabasi Marta~C.Gonzale, Cesar~A.Hidalgo.
\newblock Understanding individual human mobility patterns.
\newblock {\em Nature}, 453, June 2008.

\bibitem{trajPatMin}
Fasca Giannoti Fabio~Pinelli Micro~Nanni, Dino~Pedreschi.
\newblock Trajectory pattern mining.
\newblock {\em KDD}, pages 330--339, 2007.

\bibitem{mobisim}
MobisSim.
\newblock Project.
\newblock Available online in
  \url{http://www.masoudmoshref.com/old/myworks/documentpages/mobility_simulator.htm}.

\bibitem{comNet}
M.~E.~J. Newman.
\newblock The structure and function of complex networks.
\newblock 2003.

\bibitem{SunWSMYS10}
Jingbo Sun, Yue Wang, Hongbo Si, Xia Mao, Jian Yuan, and Xiuming Shan.
\newblock A pca-based approach for exploring space-time structure of urban
  mobility dynamics.
\newblock In Ahmed Helmy, Peter Mueller, and Yan Zhang, editors, {\em
  Proceedings of the 6th International Wireless Communications and Mobile
  Computing Conference, IWCMC 2010, Caen, France, June 28 - July 2, 2010},
  pages 859--863. ACM, 2010.

\bibitem{compNetAnHuM}
Franck~Legendre Thrasyvoulos~Spyropoulos, Theus~Hossmann.
\newblock A complex network analysis of human mobility.
\newblock {\em 3rd IEEE International Workshop on Network Science for
  Communication Networks}, April 2011.

\end{thebibliography}
              \end{multicols}
\end{document}